\documentclass[twocolumn,trackchanges]{aastex631}
\usepackage{multirow}
\usepackage{tabularx}
\usepackage{CJK}

\begin{document}

\begin{CJK*}{UTF8}{gbsn}
\title{Inferring photospheric horizontal flows from multiple observations with SUVEL models}

\correspondingauthor{Jiajia Liu}
\email{jiajialiu@ustc.edu.cn}
\author[0009-0008-8972-2726]{Quan Xie}
\affiliation{National Key Laboratory of Deep Space Exploration, School of Earth and Space Sciences, University of Science and Technology of China, Hefei 230026, China}

\author[0000-0003-2569-1840]{Jiajia Liu}
\affiliation{National Key Laboratory of Deep Space Exploration, School of Earth and Space Sciences, University of Science and Technology of China, Hefei 230026, China}
\affiliation{CAS Center for Excellence in Comparative Planetology/CAS Key Laboratory of Geospace Environment/Mengcheng National Geophysical Observatory, University of Science and Technology of China, Hefei, 230026, China}

\author[0000-0003-3439-4127]{Robert Erd{\'e}lyi}
\affiliation{Solar Physics \& Space Plasma Research Center (SP2RC), School of Mathematical and Physical Sciences\\
University of Sheffield, Hicks Building, Hounsfield Road, Sheffield S3 7RH, UK}
\affiliation{Department of Astronomy, E\"otv\"os Lor\'and University, P\'azm\'any P\'eter s\'et\'any 1/A, H-1112 Budapest, Hungary}
\affiliation{Hungarian Solar Physics Foundation, Pet\H{o}fi t\'er 3, H-5700 Gyula, Hungary}

\author[0000-0002-8887-3919]{Yuming Wang}
\affiliation{National Key Laboratory of Deep Space Exploration, School of Earth and Space Sciences, University of Science and Technology of China, Hefei 230026, China}
\affiliation{CAS Center for Excellence in Comparative Planetology/CAS Key Laboratory of Geospace Environment/Mengcheng National Geophysical Observatory, University of Science and Technology of China, Hefei, 230026, China}

\begin{abstract}
Photospheric horizontal velocity fields play essential roles in the formation and evolution of numerous solar activities. Various methods for estimating the horizontal velocity field have been proposed in the past. Aiming at the highest available (and future) spatial resolution (10 km pixel\(^{-1}\)) observations, a new method the Shallow U-net models ({\it SUVEL}) based on realistic numerical simulation and machine learning techniques was recently developed to track the photospheric horizontal velocity fields. Although {\it SUVEL} has been tested on numerical simulation data, its performance on solar observational data remained unclear. In this work, we apply {\it SUVEL} to the photospheric intensity observations from four ground-based solar telescopes (DKIST, GST, NVST, and SST) with the largest available apertures, and compare the results obtained from {\it SUVEL} with the Fourier local correlation tracking method (FLCT). Average correlation indices between granular regions and velocity fields inferred by {\it SUVEL} (FLCT) are 0.63, 0.81, 0.80, and 0.87 (0.00, 0.11, 0.16, and 0.10) for DKIST, GST, NVST, and SST observations. Higher correlation indices between the velocity fields tracked by {\it SUVEL} and granular patterns than FLCT reveal the superior performance of {\it SUVEL}, validating its reliability with respect to solar observational data.
\end{abstract}

\keywords{Sun: activity – Sun: photosphere – Sun: magnetic fields – Methods: data analysis}

\section{Introduction} \label{sec:intro}
The solar photosphere, the visible surface of the Sun, is a complex and highly dynamic system, and is filled with plasma flows. The most prominent features of the solar photosphere, including granulation, mesogranulation, and supergranulation, are all convective cells with different scales that originate in various depths in the convection zone beneath the photosphere. Horizontal flows play a key role in the formation and evolution of these cells. Moreover, abundant studies have revealed the associations between complex horizontal plasma motions and a range of solar phenomena, including small-scale photospheric vortices \citep[e.g.,][]{wang1995vorticity, wedemeyer2012magnetic, 2019ApJ...872...22L, 2019NatCo..10.3504L, 2019A&A...632A..97L, 2023A&A...674A.142L, tziotziou2023vortex, 2025ApJ...979...27X}, sunspots (active regions) \citep[e.g.,][]{1910MNRAS..70..217E, 2007A&A...468.1083Y, bi2016observation, gou2024high}, spicules and jets \citep[e.g.,][]{sterling2000solar, 2010ApJ...710.1857M, 2016ApJ...833..150L, tian2021upflows, dey2022polymeric}. Tracking the dynamic photospheric horizontal velocity fields contributes to investigations of the formation processes and mechanisms of these solar activities.

Up to now, various methods have been proposed to estimate the photospheric horizontal velocity fields. Feature tracking (FT) methods map the velocity fields by following the footprints of individual features in time-series images \citep[see e.g.,][]{1995ESASP.376b.213S}. Then, several similar methods based on local correlation tracking \citep[LCT;][]{november1988precise} are developed, including coherent structure tracking \citep[CST;][]{rieutord2007tracking}, Fourier local corelation tracking method \citep[FLCT;][]{Welsch2004, fisher2008subsurface}, differential affine velocity estimators \citep[DAVE;][]{2006ApJ...646.1358S}, and nonlinear affine velocity estimators \citep[NAVE;][]{chae2008test}. The above techniques accept the consecutive intensity images or line-of-sight (LOS) magnetograms as the input and have been applied to research from various aspects in solar physics \citep[e.g.,][]{giagkiozis2018vortex, 2023ApJS..267...35Y}. 

Recently, machine (and deep) learning techniques have achieved rapid development and promoted significant progress in multiple fields, also conducing automatic tracking of solar horizontal velocity fields. \cite{2017A&A...604A..11A} and \cite{2020FrASS...7...25T} trained two neural networks using the numerical simulation data and published two models named {\it DeepVel} and {\it DeepVelU}, respectively. Due to the low spatial and temporal resolution of the training data, the performance of models in high-resolution data still requires improvement. 

\cite{2025A&A...698A.263L} trained the shallow U-net models using high-resolution Bifrost \citep{gudiksen2011stellar, 2016A&A...585A...4C} numerical simulation data to track horizontal velocity fields from the highest available spatial resolution (\(\sim\)10 km pixel\(^{-1}\)) observations provided by the Daniel K. Inouye Solar Telescope \citep[DKIST,][]{Rimmele2020}. The models also aim to serve future telescopes, including the European Solar Telescope \citep[EST, 4-m aperture, under construction,][]{Collados2013est}, and the Chinese Giant Solar Telescope \citep[CGST, 8-m aperture, actively promoted,][]{Liu_Deng_Ji_2013}. The trained models have been integrated into a software package named {\it SUVEL}, available at the GitHub repository (\url{https://github.com/pydl/suvel}), and the version used in this work is available on Zenodo \citep{liu_2025_18082382}.

Although {\it SUVEL} has been tested and compared with FLCT on Bifrost and CO\(^5\)BOLD numerical simulation data \citep{freytag2012simulations}, it has not yet been comprehensively tested on solar observational data. In this work, we apply {\it SUVEL} to photospheric intensity images from multiple telescopes and compare the performance between {\it SUVEL} and FLCT, in order to validate the strength and advantages of {\it SUVEL}. In Section~\ref{sec:data and method}, we first briefly review {\it SUVEL} and then introduce the data used in this study. The results are presented in Section~\ref{sec:results}, followed by the whole summary of the paper and further discussions in Section~\ref{sec:cons and diss}.

\section{Methods and observations} \label{sec:data and method}
U-net, proposed by \cite{10.1007/978-3-319-24574-4_28}, is widely used for segmentation tasks for biomedical images. Its name originates from the shape of its architecture, which resembles the letter ``U''. Some examples of its details and architecture can be found in Figure 2 in \cite{2024ApJ...972..187L} and \cite{2025A&A...698A.263L}. Besides in biomedical sciences and other fields, the U-net neural network has also been employed for solar image segmentation and feature detection, e.g., the segmentation of coronal holes \citep{2018MNRAS.481.5014I}, the detection of off-limb coronal jets \citep{2024ApJ...972..187L} and sunspots \citep{2025ApJ...980..261C}, and the identification and tracking of filaments \citep{2024ApJ...965..150Z}. 

Based on a modified shallow U-net architecture, the photospheric intensity images obtained from the realistic 3D MHD simulation \citep{2016A&A...585A...4C, 2025A&A...698A.263L} were first used to train the intensity model with three consecutive intensity images as input. Considering the potential of LOS magnetograms to infer horizontal velocity fields \citep[see e.g.,][]{2020FrASS...7...25T}, the magnetic model, taking three consecutive frames of the vertical magnetic field strength, was also trained and built. Moreover, the third model, the hybrid model, takes three frames of simultaneous intensity and magnetic images, the output of the intensity model, and the output of the magnetic model as its input. Because the hybrid model learns from both photospheric intensity and vertical magnetic field strength, it was expected and then proven to perform better than the intensity and magnetic models. Therefore, {\it SUVEL}, with the architecture shown in Figure 1 in \cite{2025A&A...698A.263L}, combining these three shallow U-net models, can not only take the single photospheric intensity images or LOS magnetograms as input, but also both of them. 
 
It is worth noting that the performance of {\it SUVEL} on data with different pixel sizes and cadences was also tested. The results, shown in \cite{2025A&A...698A.263L}, indicated that {\it SUVEL} generalizes well from the training conditions (10 km pixel\(^{-1}\), 10 s cadence) to solar observations from most telescopes and satellites, covering pixel sizes of \(\sim\)10-160 km and cadences of \(\sim\)10-50 s.

To validate the performance of {\it SUVEL} on solar observational data, the data utilized in this study contain four datasets from four telescopes, respectively, including DKIST, the 1.6 m Goode Solar Telescope \citep[GST;][]{cao2010scientific}, the New Vacuum Solar Telescope \citep[NVST;][]{liu2014new}, and the Swedish 1-m Solar Telescope \citep[SST;][]{scharmer20031}. Relevant information of the observations and properties of the corresponding data are summarized in Table~\ref{Tab:obs}. 

\begin{figure*}[ht!]
    \centering
    \includegraphics[width=1.0\textwidth]{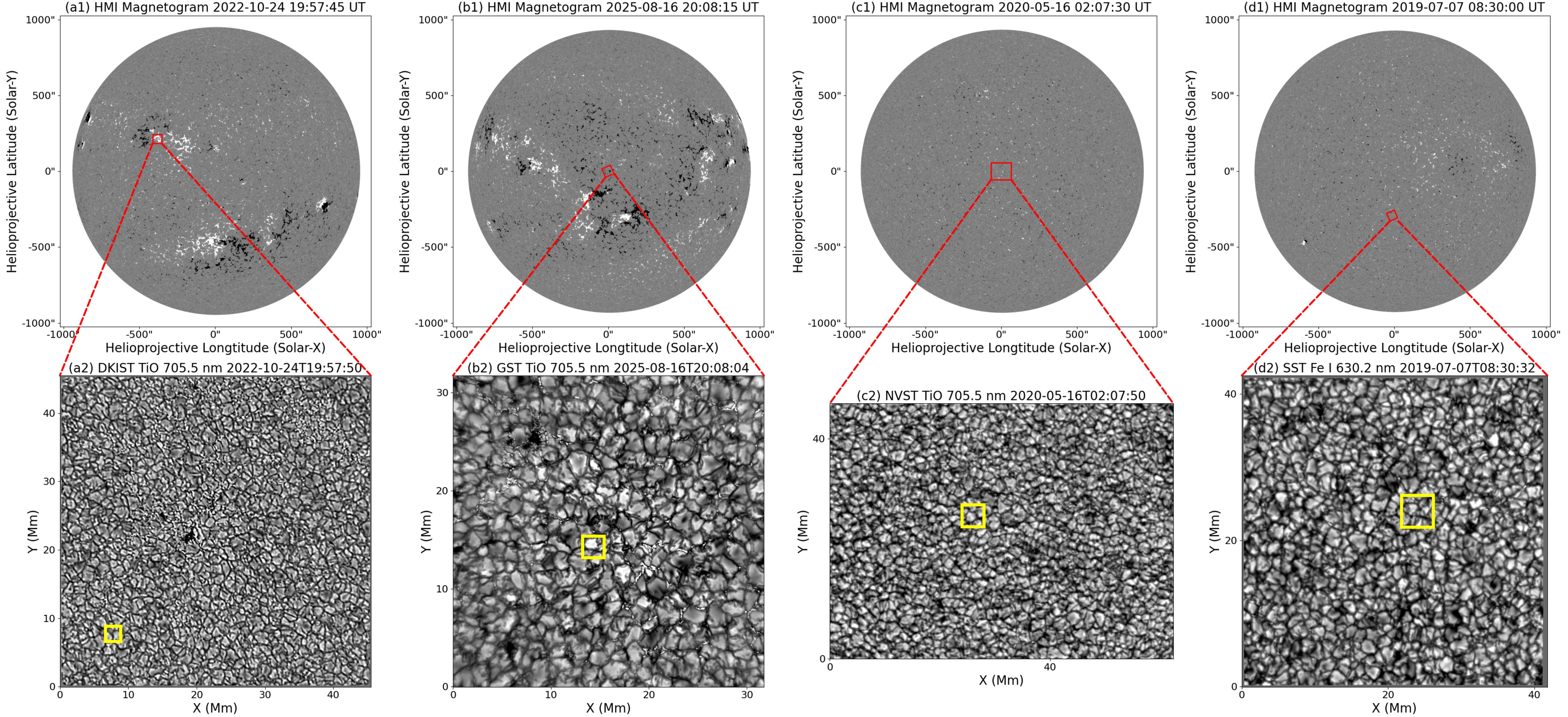}
    \caption{The four red squares drawn on the SDO/HMI magnetograms in panels (a1)-(d1) depict the FOV locations of DKIST, GST, NVST, and SST observations, respectively. The bottom row shows four example intensity images of an instance in the four observations. The yellow squares outline the regions for further study.}
    \label{fig_obs}
\end{figure*}

All four observations are from the visible light band (TiO 705.7 nm wideband and Fe I 630.2 nm wideband), focusing on the photospheric surface layer. The cadences and pixel sizes are all suitable for {\it SUVEL}. The locations of field-of-view (FOV) are plotted with red squares in the Helioseismic Magnetic Imager \citep[HMI;][]{scherrer2012helioseismic} magnetograms, shown in the top row in Figure~\ref{fig_obs}, respectively. 

The FOVs of SST and GST observations are tilted 70\(^\circ\) clockwise and 23.2\(^\circ\) counterclockwise relative to the Sun's north pole, while the DKIST and NVST FOVs have no obvious rotations. Observations from GST, NVST, and SST target the quiet sun (QS) region, while the DKIST observations focus on an active region (AR). The bottom row presents four example intensity images from the datasets, captured by DKIST, GST, NVST, and SST, respectively. It is worth noting that we removed some empty areas near the boundary from the GST and NVST in this study. That is the reason why the sizes of the intensity images from GST and NVST (Figure~\ref{fig_obs}b2 and c2) are smaller than the FOV size shown in Table~\ref{Tab:obs}. 

Following the procedure from \cite{2025A&A...698A.263L}, the intensity images of all four datasets are firstly unified, respectively. For each frame, the image is divided by its mean intensity and then multiplied by the overall mean intensity of the dataset. It ensures that all images of the four datasets have the same average intensity, respectively. Then, these unified intensity data are normalized to [0, 1] and input to the intensity model of {\it SUVEL}. The model output, velocity along the \(x\)-axis (\(v_{xi}\)) and velocity along the \(y\)-axis (\(v_{yi}\)) are re-scaled to the real velocity (\(v_{xir}\) and \(v_{yir}\)) in units of km/s, based on the Equation 6 in \cite{2025A&A...698A.263L} as follows:
\begin{equation}
    \begin{aligned}
        v_{xir} = v_{xi} \cdot 23.4 - 11.7, \quad
        v_{yir} = v_{yi} \cdot 23.4 - 11.7.
    \end{aligned}
    \label{eq_transf}
\end{equation}

The above unified and normalized intensity data are also utilized by FLCT to estimate the velocity fields. For the FLCT method utilized in this study, the local horizontal velocity field is obtained by maximizing the spatially localized cross-correlation between two consecutive intensity images. Following \cite{november1988precise}, let \(J_t(\mathbf{x})\) and \(J_{t+\tau}(\mathbf{x})\) denote the photospheric intensity images at times \(t\) and \(t+\tau\), where the sampling time delay \(\tau\) is also the cadence of the observation. The cross-correlation function at location \(\mathbf{x}\) for a displacement vector \(\boldsymbol{\delta}\) is defined as:
\begin{equation}
C(\boldsymbol{\delta}, \mathbf{x}) = 
\int J_t\!\left(\boldsymbol{\xi} - \frac{\boldsymbol{\delta}}{2}\right)
J_{t+\tau}\!\left(\boldsymbol{\xi} + \frac{\boldsymbol{\delta}}{2}\right)
W_\sigma(\mathbf{x} - \boldsymbol{\xi}) \, d^2\boldsymbol{\xi}.
\label{eq:flct_corr}
\end{equation}

Here, $W_\sigma$ is an apodizing window that confines the correlation to a local neighborhood. Same as \citet{fisher2008subsurface}, we adopt a two-dimensional Gaussian window:
\begin{equation}
W_\sigma(\mathbf{s}) =  \exp\!\left( -\frac{|\mathbf{s}|^2}{\sigma^2} \right),
\label{eq:gaussian_window}
\end{equation}
where $\mathbf{s} = \mathbf{x} - \boldsymbol{\xi}$ is the relative displacement from the integration point $\boldsymbol{\xi}$ to the window center $\mathbf{x}$, and $\sigma$ is the standard deviation of the Gaussian, serving as the fundamental spatial scale parameter in FLCT.

The parameters of FLCT are similar to the previous studies \citep[see e.g.,][]{2019ApJ...872...22L, 2019NatCo..10.3504L, 2025ApJ...979...27X}. However, it is worth noting that the Gaussian filter's \(\sigma\) is set specifically for each dataset to avoid some pixels moving outside the window. Assuming the maximum photospheric horizontal speed is 10 km s\(^{-1}\), the maximum movement of a pixel is 10 km s\(^{-1}\) \(\cdot\) dt / ds, with ds and dt representing the pixel size and cadence of the observation. Because a pixel can move in all directions, the value of sigma should be set twice the maximum movement. Therefore, the values of sigma are set 15, 15, 15, and 10 for DKIST, GST, NVST, and SST datasets, respectively.

\section{Results} \label{sec:results}
\subsection{Granules and velocity fields}
The solar granulation pattern, first observed and described by \cite{1801RSPT...91..265H}, was initially interpreted as arising from ``hot clouds'' floating above a cooler solar surface. Later, \cite{1864MNRAS..24..161D} coined the term granules to describe the bright, irregularly shaped polygonal cells interspersed with narrow dark lanes. Subsequent Doppler shift measurements indicated localized upward motions within the granular regions (hereafter GR) and downward motions within the intergranular lanes (hereafter IGL). Thus, the dynamics of granulation can be understood as hot plasma rising in the GR and cooler plasma descending in the IGL. Consequently, velocity fields are expected to diverge in the GR and converge within the IGL.

\begin{figure*}[ht!]
    \centering
    \includegraphics[width=1.0\textwidth, height=0.9\textheight]{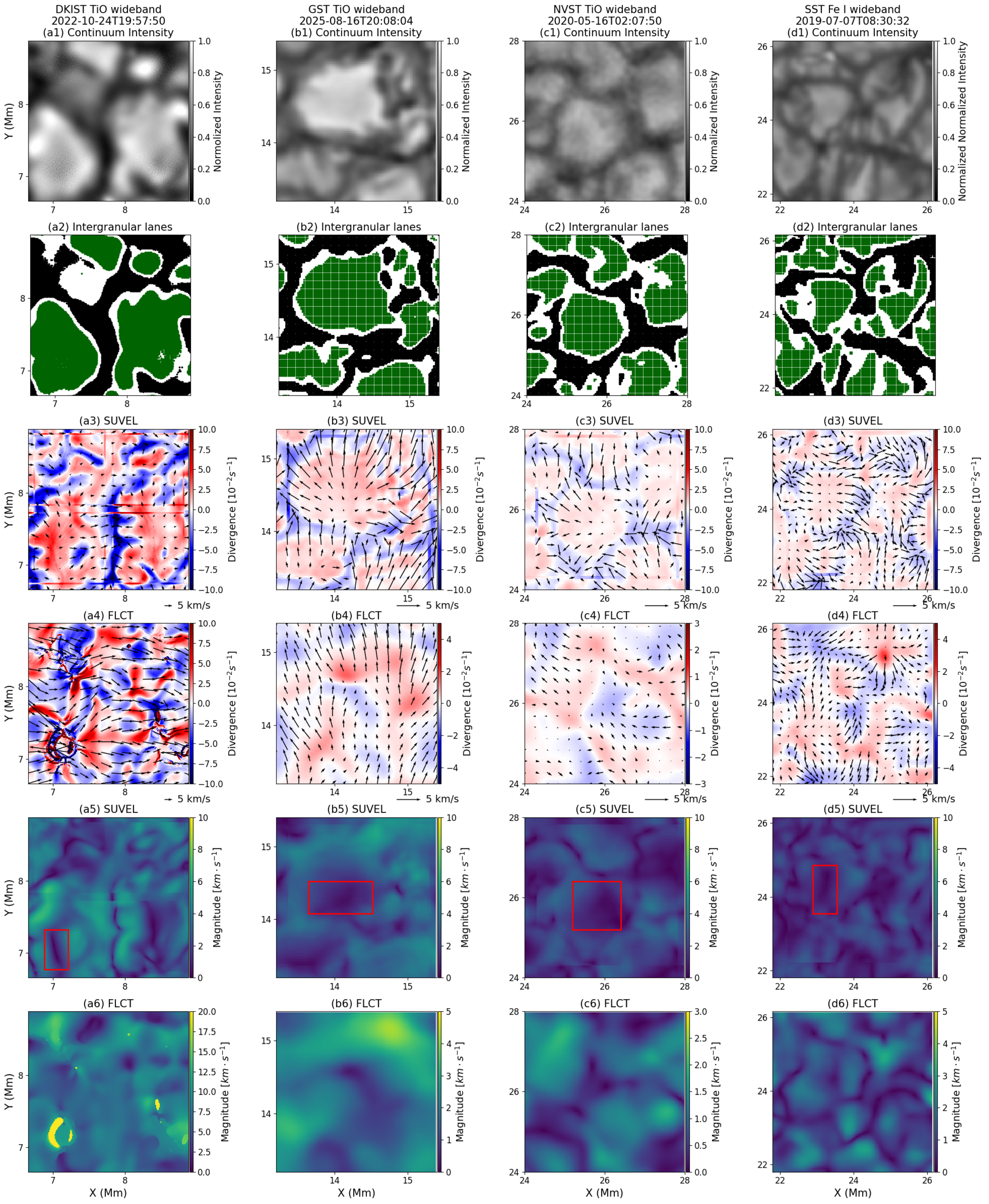}
    \caption{(a1)-(d1) Close-up views of the yellow boxes shown in Figure~\ref{fig_obs}. (a2)-(d2) Green and black dots represent the granules regions and intergranular lanes. (a3)-(d3) The black arrows depict the velocity fields inferred by {\it SUVEL}, with the backgrounds showing the corresponding divergence maps. (a4)-(d4) Similar to (a3)-(d3), but for the velocity fields estimated by FLCT. (a5)-(d5) The magnitude maps of velocity fields inferred by {\it SUVEL}. (a6)-(d6) Similar to (a5)-(d5), but for the results of FLCT.}
    \label{fig_sample}
\end{figure*}

We selected four small subregions from the four observational datasets, with their locations marked by yellow squares in Figure~\ref{fig_obs}(a2)-(d2), to showcase the correlations between GR and IGL with the velocity fields derived from {\it SUVEL} and FLCT. Representative intensity images of the sampled regions are shown in the first row of Figure~\ref{fig_sample}. 

Several granules and the surrounding intergranular lanes are clearly visible in these images (Figures~\ref{fig_sample}a1-d1). In this study, pixels with intensities higher than \(\mu\) + \(\sigma\) are identified as GR, while those with intensities lower than \(\mu\) - \(\sigma\) are classified as IGL, where \(\mu\) and \(\sigma\) denote the mean and standard deviation of the image intensity, respectively. The GR and IGL pixels are indicated by green and black dots in Figures~\ref{fig_sample}(a2)-(d2). Black arrows in Figures~\ref{fig_sample}(a3)-(d3) and (a4)-(d4) represent the velocity fields obtained from {\it SUVEL} and FLCT, respectively, overlaid on the corresponding divergence maps. For velocity along the \(x\)-axis (\(v_{x}\)) and velocity along the \(y\)-axis (\(v_{y})\), the corresponding divergence map is calculated using the below formula:
\begin{equation}
\nabla \cdot \mathbf{v} =
\left( \hat{\mathbf{i}}\frac{\partial}{\partial x} + \hat{\mathbf{j}}\frac{\partial}{\partial y} \right)
\cdot
\left( v_x \hat{\mathbf{i}} + v_y \hat{\mathbf{j}} \right)
= \frac{\partial v_x}{\partial x} + \frac{\partial v_y}{\partial y}.
\label{Equ:div}
\end{equation}
Positive value means divergence, while negative one represents convergence. Furthermore, Figures~\ref{fig_sample}(a5)-(d5) and (a6)-(d6) show the velocity magnitude maps estimated by {\it SUVEL} and FLCT, respectively.

By comparing the divergence maps derived from {\it SUVEL} and FLCT with the corresponding intensity images, it is evident that the velocity fields tracked by {\it SUVEL} in all four observations are more consistent with the distributions of granules (GR) and intergranular lanes (IGL): exhibiting divergence in GR and convergence in IGL. In contrast, the performance of FLCT appears clearly unsatisfactory across all four observations. The correlation between the velocity fields obtained by FLCT and the spatial locations of GR and IGL is weak. To quantitatively assess this relationship, we propose a method, motivated by the approach of \cite{2019NatCo..10.3504L}, to estimate the correlation between the velocity fields and the corresponding intensity structures.

Firstly, based on the points of GR and IGL shown in Figure~\ref{fig_sample}(a2)-(d2), we set all GR points to one and all other points to zero, generating the GR binary intensity map I\(_1\). Similarly, for the IGL binary intensity map I\(_2\), all IGL points are set as -1, with other points set as zero. The total number of all GR and IGL points are {\it N\(_{G}\)} and {\it N\(_{IG}\)}.

Then, divergence maps D are calculated from the horizontal maps estimated by {\it SUVEL} and FLCT, using Equation (\ref{Equ:div}). Furthermore, the sign map D\(_s\) of the divergence map D is obtained, meaning all positive values in D are set as 1 and all negative values are set as -1. 

Lastly, we combine I\(_1\)(I\(_2\)) and D\(_s\) through a Hadamard product, and obtain the correlation map \(C_1\) (\(C_2\)) of GR (and IGL), respectively. Correlation indices CI\(_1\) (and CI\(_2\)) between GR (and IGL) and velocity fields are calculated correspondingly with the following equation:
\begin{equation}
    \mathrm{CI}_1 = (\sum C_1) / N_G,  \quad \mathrm{CI}_2  = (\sum C_2) / N_{IG}.
\end{equation}
The above procedures indicate that CI are concentrated on [-1, 1], with higher values implying higher relations. Correlation indices (CI\(_1\) and CI\(_2\)) of all frames in the four observations are calculated, using the velocity fields estimated by {\it SUVEL} and FLCT, respectively. 

\begin{figure*}[ht!]
    \centering
    \includegraphics[width=1.0\textwidth]{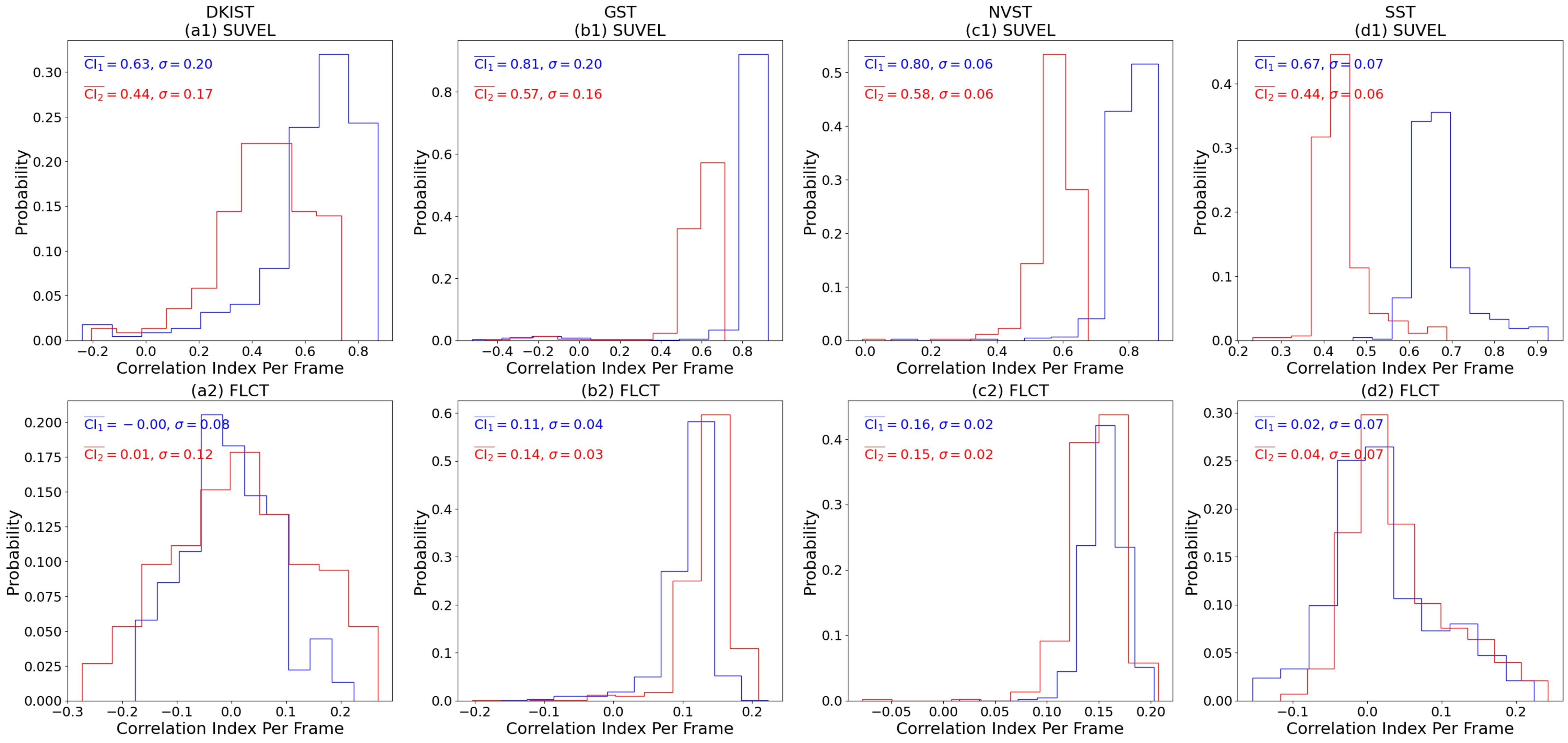}
    \caption{Per-frame average correlation indices between velocity fields and granulation patterns. CI\(_1\) (blue) and CI\(_2\) (red) depict the results of granule regions and intergranular lanes, respectively. (a1)-(d1) represent the results of the velocity fields inferred by {\it SUVEL}. (a2)-(d2) depict the results from FLCT.}
    \label{fig_CI}
\end{figure*}

Shown in Figure~\ref{fig_CI}, statistical results indicate that the average CIs between GR and velocity fields inferred by {\it SUVEL} are 0.63, 0.81, 0.80, and 0.87 for DKIST, GST, NVST, and SST observations, all higher than the results (0.00, 0.11, 0.16, and 0.10) from FLCT. A similar priority also holds for IGL, suggesting that {\it SUVEL} provides better and more reliable photospheric horizontal velocity field estimation than FLCT.

Comparing the distributions of CI\(_1\) and CI\(_2\) from {\it SUVEL}, shown in Figure~\ref{fig_CI}(a1)-(d1), it is worth noting that {\it SUVEL} tracks velocity fields less effectively in intergranular lanes than in granules. We suggest that one of the reasons is the intrinsic limitations of {\it SUVEL}. \cite{2025A&A...698A.263L} noted that {\it SUVEL} still exhibits certain shortcomings and requires further improvement, particularly when applied to regions rich in small-scale structures, due to the nature of the U-NET neural network architecture. Intergranular lanes and their surrounding regions are where small-scale structures are densely concentrated \citep[e.g.,][]{2009LRSP....6....2N}, which influences the velocity field reconstruction of {\it SUVEL}. A key point of future work is to focus on the improvement of {\it SUVEL} on such small-scale structures.

Next, let us also apply the method to the Bifrost numerical simulation data \citep[details see][]{2025A&A...698A.263L}, which was used for the training and testing of {\it SUVEL} models. Instead of applying estimated velocity fields, we adopt simulated intensity images and their corresponding velocity fields to calculate the average CIs for GR and IGL. CIs of GR and IGL are 0.70 \(\texttt{$\pm$}\) 0.05 and 0.51 \(\texttt{$\pm$}\) 0.03, respectively. These results are close to the results of {\it SUVEL} obtained from the above various observations, again, supporting the reliability of {\it SUVEL}.  

Moreover, focusing on the magnitude distribution of the velocity fields derived by {\it SUVEL}, one can notice that there is usually a small region within a granule with low horizontal speeds. The red squares in panels (a5)-(d5) in Figure~\ref{fig_sample} mark four example positions. Naturally, a cloud of cool plasma arises onto the surface from the beneath layer, experiencing low horizontal speeds corresponding to the red squares in Figure~\ref{fig_sample}(a5)-(d5). The plasma then spreads out, contributing to the divergent flows inside GRs. The plasma moves faster and faster, reaching its peak speed in the IGL. Bunches of rapid plasma flows from neighbouring granules gather together and begin to sink due to their relatively low temperature. This also naturally causes the relatively low horizontal velocities at the IGLs, which can be seen when comparing panels (a3)-(d3) and panels (a5)-(d5) in Figure~\ref{fig_sample}.

Let us now turn our attention to the velocity fields estimated by FLCT, shown in panels (a4)-(d4) and (a6)-(d6) in Figure~\ref{fig_sample}. Although FLCT can also capture some details, most structures, especially the boundaries between GRs and IGLs, disappear. \cite{2013A&A...555A.136V} used CO\(^5\)BOLD numerical simulation data to evaluate LCT, suggesting that LCT underestimates the speed by a factor of approximately three. Comparisons between velocity fields, derived by {\it SUVEL} and FLCT from GST, NVST, and SST observations, also indicate a similar underestimation of velocity fields estimated by FLCT as {\it SUVEL}, shown in Figures~\ref{fig_sample}(b5)-(d5) and (b6)-(d6). This finding further proves the accuracy of {\it SUVEL} and supports the previous findings using observational evidence. However, for DKIST observations, it is seen that FLCT infers larger speeds than {\it SUVEL}, shown in Figures~\ref{fig_sample}(a5) and (a6). These significantly larger speeds (exceeding 20 km \(\cdot\) s\(^{-1}\) in the yellow regions in Fig.~\ref{fig_sample}a6) should not be real, given that the sound speed and Alfv{\'e}n speed in the photosphere are generally less than 10 km \(\cdot\) s\(^{-1}\). These unrealistic large speeds may be attributed to the high spatial resolution (11.1 km pixel\(^{-1}\)), which influences the performance of FLCT.

In summary, based on the above results and analysis, we conclude that {\it SUVEL} models can credibly reconstruct the photospheric horizontal velocity fields from the observations of different telescopes, proving its usability on observational data. 

\subsection{Influence on vortex detections}
Ubiquitous small-scale velocity/plasma vortices are found in the photosphere, which are primarily located in the intergranular lanes \citep[see e.g.,][]{giagkiozis2018vortex, 2019ApJ...872...22L, tziotziou2023vortex}. Recently, \cite{2025ApJ...979...27X} found that many small-scale vortices tend to be located at larger scales in inter-mesogranular lanes, offering a new perspective on the depth at which such small-scale vortices form beneath the solar photosphere. Photospheric vortices were generally believed to be associated with coronal heating \citep[e.g.,][]{1993SoPh..147...47Z}. \cite{2019NatCo..10.3504L} showed that vortices transport energy from the photosphere to the upper chromosphere. Note that these small-scale vortices are not easy to identify with eyes. Various methods have been developed to reduce the biases and limitations inherent in manual detection. For example, based on the \(\Gamma\)-functions method proposed by \cite{graftieaux2001combining}, \cite{2019ApJ...872...22L} developed and made it openly available the automated swirl detection algorithm (ASDA) to identify vortices from estimated photospheric horizontal velocity fields. The algorithm (and the \(\Gamma\)-functions method) was recently improved by \cite{2025A&A...700A...6X}, named as the Optimized ASDA and available at the GitHub repository (\url{https://github.com/dreamstar0831/Optimized_ASDA}), resulting in improved detection rates and accuracies of such small-scale vortices.

Previous studies are mostly dependent on the velocity fields derived from FLCT \citep[e.g.,][]{giagkiozis2018vortex, 2019ApJ...872...22L, 2019NatCo..10.3504L, 2025ApJ...979...27X}. However, as mentioned before, the performances of FLCT seem not as precise as expected. \cite{tziotziou2023vortex} also pointed out that very small photospheric vortices might be ignored by LCT methods. Therefore, to explore how different velocity tracking methods ({\it SUVEL} and FLCT) influence the vortex detection, we employ the Optimized ASDA with velocity fields tracked by {\it SUVEL} and FLCT from the DKIST, GST, NVST, and SST observations, introduced in Section~\ref{sec:data and method}.

\begin{figure*}[ht!]
    \centering
    \includegraphics[width=1.0\textwidth]{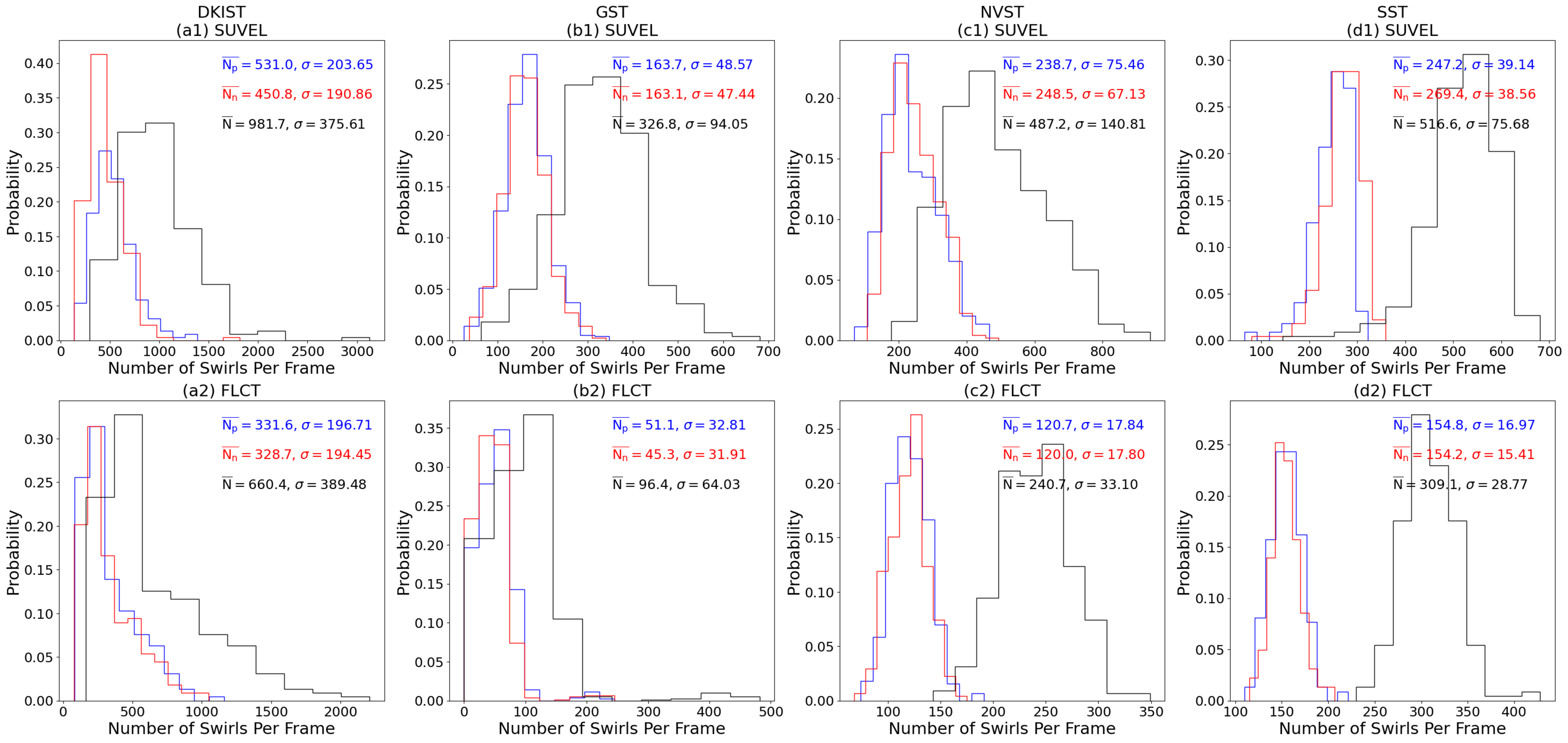}
    \caption{Per-frame average number of detected vortices. Subscripts p and n (blue and red) denote positive and negative vortices, respectively. Black curves and texts are the results of all swirls. \(\sigma\) represents the standard deviation. (a1)-(d1) Number of vortices detected by the Optimized ASDA and {\it SUVEL} from the DKIST, GST, NVST, and SST observations. (a2)-(d2) Similar to the top panels, but for the results of FLCT.}
    \label{fig_num}
\end{figure*}

Figure~\ref{fig_num} depicts the average number of vortices detected per frame using {\it SUVEL} and FLCT with the Optimized ASDA from DKIST, GST, NVST, and SST observations, respectively. \(N_p\) and \(N_n\) denote the number of positive (counterclockwise rotation) and negative (clockwise rotation) vortices, distinguished by blue and red, respectively. The average number of all vortices per frame is represented with \(N\) in black. We note that vortices, detected from DKIST intensity images, whose radii are less than 5 pixels, are omitted, because the diffraction limit of the TiO 705 nm band for DKIST is estimated approximately 0.038\(^{\prime\prime}\) (2.5 pixels). The number densities of vortices are found to be 0.47, 0.32, 0.17, and 0.29 Mm\(^{-2}\) using {\it SUVEL} on the observations from DKIST, GST, NVST, and SST, respectively. On the other hand, using FLCT, the number densities are 0.32, 0.09, 0.08, and 0.17 Mm\(^{-2}\). Considering the different spatial resolutions of the four telescopes, it is found that the number densities are nearly positively related to the pixel sizes of observations. \cite{2020ApJ...894L..17Y} pointed out that larger vortices detected from low spatial resolution observations may consist of several smaller vortices detected from high-resolution data, explaining why more vortices are detected from high-resolution data. This is consistent with the above findings that the most significant number densities of vortices are found with DKIST observations no matter which horizontal velocity field tracking method has been used.

Comparing the vortex numbers estimated by {\it SUVEL} and FLCT, it is evident that the vortex numbers from FLCT are all lower than those from {\it SUVEL} for all four observations. Therefore, FLCT might lead to an underestimation of vortex number (density). Moreover, the densities from the four datasets are approximately 0.3 Mm\(^{-2}\), significantly greater than almost all previous results, e.g., 2.2\(\times\)10\(^{-2}\) Mm\(^{-1}\) \citep{2010ApJ...723L.139B}, 6.1\(\times\)10\(^{-2}\) Mm\(^{-1}\) \citep{2019NatCo..10.3504L}, and 2.11\(\times\)10\(^{-2}\) Mm\(^{-1}\) \citep{2025ApJ...979...27X}. In addition to {\it SUVEL}, another factor contributing to the more detection is that the Optimized ASDA could miss fewer vortices than the original ASDA, as suggested in \cite{2025A&A...700A...6X}. 

We also investigated the distributions of vortex radius, rotation speed, and expansion/contraction speed derived from the four observations using {\it SUVEL}. Vortices with smaller radii are detected in higher-resolution observations, suggesting that even smaller vortices could exist in the solar atmosphere. Rotation speeds are all found around 2 km s\(^{-1}\), which are consistent with some previous studies \citep[e.g.,][]{2009A&A...507L...9W}, especially with our previous results from numerical simulations \citep{2019A&A...632A..97L}, but higher than those found with FLCT \citep{2019ApJ...872...22L}, shedding light on the real rotation speed of photospheric vortices \citep{2024A&A...682A.181C}. Expansion (contraction) speeds are approximately 0.5 (-0.6) km s\(^{-1}\), again, larger than \(\sim\) 0.2 (-0.2) km s\(^{-1}\) in previous studies \citep{2019ApJ...872...22L, 2019NatCo..10.3504L, 2025ApJ...979...27X}. Therefore, combining {\it SUVEL} with the Optimized ASDA enables more vortices to be detected with higher accuracy in their positions, radii, expansion speeds, and other properties, contributing to further detailed studies of photospheric vortices.

\section{Conclusion and discussions} \label{sec:cons and diss}

In this study, we applied the newly developed horizontal velocity field estimation model {\it SUVEL} to solar observations from multiple telescopes (DKIST, GST, NVST, and SST), to validate the strength and advantages of {\it SUVEL}. The main results of this work are summarized as follows. 
\begin{enumerate}
    \item By sampling four small regions of an instance from the four observations, we intuitively discover that velocity fields derived by {\it SUVEL} match the corresponding granular pattern better than the ones estimated by FLCT. We then propose a method to calculate the correlation indices between the velocity fields and intensity structures. Higher correlation indices of all four observations further prove the priority of {\it SUVEL}. 
    
    \item We combined the Optimized Automated Swirl Detection Algorithm (Optimized ASDA) with {\it SUVEL} and FLCT to detect photospheric vortices. More vortices are identified using {\it SUVEL} than using FLCT, indicating that FLCT underestimates the vortex number in previous studies. The open-source software {\it SUVEL} can provide better and reliable results for public use. It is worth noting that the number of true vortices detected by {\it SUVEL} and FLCT is still unclear. Future work focusing on the one-to-one comparisons between vortices detected from {\it SUVEL} (FLCT) and the real vortices identified from the simulated velocity fields will further reveal the accuracy of  {\it SUVEL} and FLCT on photospheric horizontal velocity field reconstruction. 
\end{enumerate}

Larger vortex densities also indicate the great potential of combining {\it SUVEL} with the Optimized ASDA for photospheric vortex detection. Although the influence of {\it SUVEL} and the Optimized ASDA on vortex detection has been discussed in Section~\ref{sec:results}, the influence on the estimation of vortex lifetime is still under exploration. \cite{2019ApJ...872...22L} proposed a method to estimate the lifetime and found that most vortices have lifetimes less than twice the cadences. We speculate that the results may be affected by the inaccurate velocity fields estimated by FLCT. In future work, we will also focus on the study of vortex lifetime using {\it SUVEL} and the Optimized ASDA to provide more precise lifetime distributions of photospheric vortices.

\section*{Acknowledgements}
J.L. and Q.X. acknowledge the support from the Strategic Priority Research Program of the Chinese Academy of Science (Grant No. XDB0560000), and the National Natural Science Foundation (NSFC 42188101, 12373056). R.E. acknowledges the NKFIH (OTKA, grant No. K142987) Hungary for enabling this research. R.E. is also grateful to the Science and Technology Facilities Council (STFC, grant No. ST/M000826/1) UK, PIFI (China, grant No. 2024PVA0043), and the NKFIH Excellence Grant TKP2021-NKTA-64 (Hungary). We acknowledge the use of Swedish 1-m Solar Telescope (SST) data. SST is operated on the island of La Palma (Spain) by the Institute for Solar Physics of Stockholm University in the Spanish Observatorio del Roque de los Muchachos of the Instituto de Astrof\'isica de Canarias. The research reported herein is based in part on data collected with the Daniel K. Inouye Solar Telescope (DKIST), a facility of the U.S. National Science Foundation. The National Solar Observatory operates DKIST under a cooperative agreement with the Association of Universities for Research in Astronomy, Inc. DKIST is located on land of spiritual and cultural significance to Native Hawaiian people. The use of this important site to further scientific knowledge is done with appreciation and respect.

\bibliography{sample631}{}
\bibliographystyle{aasjournal}

\begin{table*}[ht]
\centering
\caption{Summary of the four observations and datasets used in this work}
\label{Tab:obs}
\begin{tabular}{ccccccccc}
\toprule
Obs No. & Telescope & Wavelength & Time Range & FOV Center & Region & Pixel Size & Cadence & FOV Size \\
\midrule
\multirow{2}{*}{1} & DKIST & TiO (705 nm) & 2022-10-24 & ($-397^{\prime\prime}$, $198^{\prime\prime}$) & AR & 0.015$^{\prime\prime}$ (11.1 km) & 6.1 s & 45 Mm $\times$ 45 Mm \\
 & /VBI &  & T19:47$\sim$20:10 &  &  &  &  &  \\
\midrule
\multirow{2}{*}{2} & GST & TiO (705 nm) & 2025-08-16 & ($0^{\prime\prime}$, $0^{\prime\prime}$) & QS & 0.03$^{\prime\prime}$ (22.2 km) & 13 s & 45 Mm $\times$ 45 Mm \\
 &  &  & T19:27$\sim$22:17 &  &  &  &  &  \\
\midrule
\multirow{2}{*}{3} & NVST & TiO (705 nm) & 2020-05-15 & ($0^{\prime\prime}$, $0^{\prime\prime}$) & QS & 0.052$^{\prime\prime}$ (40 km) & 30 s & 102 Mm $\times$ 86 Mm \\
 &  &  & T01:17$\sim$05:02 &  &  &  &  &  \\
\midrule
\multirow{2}{*}{4} & SST & Fe I (630.2 nm) & 2019-07-07 & ($0^{\prime\prime}$, $-300^{\prime\prime}$) & QS & 0.059$^{\prime\prime}$ (43.6 km) & 4.2 s & 41 Mm $\times$ 42 Mm \\
 & /CRISP &  & T08:23$\sim$08:39 &  &  &  &  &  \\
\bottomrule
\end{tabular}
\raggedright
Notes: FOV refers to field-of-view.\\
AR refers to active region.\\
QS refers to quiet sun.
\end{table*}

\clearpage
\end{CJK*}
\end{document}